\documentclass[preprint,psfig]{aastex}

\def\ltsima{$\; \buildrel < \over \sim \;$}

\def\lsim{\lower.5ex\hbox{\ltsima}}
\def\gtsima{$\; \buildrel > \over \sim \;$}
\def\gsim{\lower.5ex\hbox{\gtsima}}

\begin{document}
\title{Measuring the Size of Quasar Broad-Line Clouds Through \\
Time Delay Light-Curve Anomalies of Gravitational Lenses}

\author{J. Stuart B. Wyithe\altaffilmark{1} and Abraham Loeb}
\affil{Harvard-Smithsonian Center for Astrophysics, 60 Garden St.,
Cambridge, MA 02138;\\
swyithe@cfa.harvard.edu, aloeb@cfa.harvard.edu}

\altaffiltext{1}{Hubble Fellow}

\begin{abstract}

Intensive monitoring campaigns have recently attempted to measure the time
delays between multiple images of gravitational lenses.  Some of the 
resulting light-curves show puzzling low-level, rapid variability which is
unique to individual images, superimposed on top of (and concurrent with)
longer time-scale intrinsic quasar variations which repeat in all images. 
We demonstrate that
both the amplitude and variability time-scale of the rapid light-curve
anomalies, as well as the correlation observed 
between intrinsic and microlensed variability, are naturally explained by 
stellar microlensing
of a smooth accretion disk which is occulted by optically-thick broad-line
clouds.  The rapid time-scale is caused by the high velocities of the
clouds ($\sim 5\times 10^3~{\rm km~s^{-1}}$), and the low amplitude results
from the large number of clouds covering the magnified or demagnified parts
of the disk. The observed amplitudes of variations in specific lenses
implies that the number of broad-line clouds that cover $\sim 10\%$ of the
quasar sky is $\sim10^5$ per $4\pi$ steradian. This is comparable to
the expected number of broad line clouds in models where the clouds originate
from bloated stars.

\end{abstract}


\section{Introduction}

The use of time delays between the images produced by galaxy scale
gravitational lenses to measure the Hubble constant was
proposed almost four decades ago (Refsdal~1964). With approximately ten
time delays now measured for different lenses (e.g. Kundic et al.~1997; 
Schechter et al.~1997; Burud et al.~2000), this method has recently become 
practical (see Kochanek~2002 for a recent summary and analysis). As part of 
the associated observational effort, Burud~(2002) has presented light-curves
with very well determined time delays for the systems RX J0911+05 and SBS
1520+530. In both systems Burud~(2002) finds evidence for short-term
variability which is unique to individual images. The variability is
observed on time-scales of tens to hundreds of days, with an amplitude of 
up to a few percent, and appears to be associated with small intrinsic 
fluctuations.  Similar rapid, low amplitude residual variability has 
previously been observed between the images of Q0957+561 (Schild~1996). 
If caused by microlensing
within the lens galaxy, this short-term variability is puzzling because
naively, the observed amplitudes require very small (non-stellar) microlens
masses (Schild~1996) of $m\la10^{-4}M_{\odot}$ (see also Gould \&
Miralda-Escud\'{e}~1997). Having a dominant microlens population in the
required mass range is ruled out by the work of Refsdal et al.~(2000) and
Wambsganss et al.~(2000) on quasar microlensing in Q0957+561 and by
Wyithe, Webster \& Turner~(2000a,b) in Q2237+0305, as well as by Galactic
microlensing experiments (e.g. Alcock et al.~2001).  Another explanation
was suggested by Gould \& Miralda-Escud\'{e}~(1997); in their model rapidly
moving hot spots (or cold spots) on the disk surface possess a high
transverse velocity and lead to short time-scale variability as the spots
move across the stellar microlensing caustics. Since only a small fraction
of the source area is magnified, the variability amplitude is small. By
postulating the existence of spots with appropriate properties, one might
avoid the need to invoke a population of planetary mass microlenses.

In this paper we offer a natural explanation for the observed properties of the 
short-term variability involving quasar phenomena identified through separate
lines of inquiry. We start our discussion with a description of the standard
quasar (\S \ref{quasar}) and microlensing (\S \ref{mlens}) models.  In \S
\ref{BLR}, we explain how the rapid, low-amplitude microlensing observed by
Burud~(2002) is naturally produced without postulating hypothetical
components for either the quasar disk or the lens population. Our model
includes typical stellar-mass microlenses, and a featureless accretion disk
surrounded by a shell of optically thick broad line clouds, 
which possess the high
velocities and small source sizes necessary to explain the rapid, low
amplitude variability. In \S \ref{spots} and \S \ref{lcurve} we compare
this model against alternative explanations. Finally, we summarize our
primary conclusions in \S~\ref{discussion}. Throughout the paper we assume
a flat (filled--beam) cosmology having density parameters of $\Omega_m=0.3$
in matter, $\Omega_\Lambda=0.7$ in a cosmological constant, and a Hubble
constant $H_0=65{\rm \,km\,sec^{-1}\,Mpc^{-1}}$.

\section{Quasar Model}
\label{quasar}

The UV-optical spectra of nearly all quasars show broad emission lines
with Doppler widths of $\sim5000\,{\rm km\,sec^{-1}}$.  The emission is
believed to originate from dense clouds having a small filling factor of
$\sim10^{-6}$, which are illuminated by a central ionizing continuum source
(see, e.g. Netzer~1990 and references therein). Two popular models were
proposed for the origin of the broad line emitters: (i) cool clouds
confined within a hot medium (Krolik, McKee \& Tarter~1981); and (ii) winds
from giant stars which are bloated due their exposure to the intense quasar
radiation (Alexander \& Netzer~1997 and references therein). The
equivalent width of the lines indicates that $\sim10\%$ of the continuum
emission of the quasar is absorbed and reprocessed by the clouds. Hence,
the fraction of sky covered by clouds when viewed from the central
continuum source, is $F_{\rm c}\sim0.1$.

The number of clouds, which is important for our analysis is the subject of 
some controversy. An upper limit for the largest number of individual 
clouds can be obtained
from photo-ionization arguments, yielding $N_{\rm cl}\la10^{7-8}$ (Arav
et al.~1998). In addition, a lower limit on the number of clouds $N_{\rm
cl}\ga10^{7-8}$, has been derived based on the smoothness of the emission
lines of NGC 4151 (Arav et al.~1998) and 3C273 (Dietrich et al.~1999),
under the assumption that the clouds are confined with a thermal internal
velocity dispersion of $\sim 10$--$20~{\rm km~s^{-1}}$. However the estimated
number of clouds is expected to go down dramatically if this assumption is
relaxed, bringing the required number to be as small as few times $10^5$
for a velocity spread of $\sim 100~{\rm km~s^{-1}}$ per cloud (H. Netzer,
private communication), as expected for the cometary tails of
bloated stars. Hence, the observed smoothness of the emission lines does
not rule out the bloated star model, which predicts the existence of
$10^{4-5}$ clouds per quasar.  It was also suggested that the low
excitation lines on which the above analysis was based, originate from the
outer parts of the accretion disk while the high excitation lines are
dominated by discrete emitters (see Dietrich et al.~1999 and references
therein). We note that the number of disk-transiting clouds on which we
focus in this paper could in principle be significantly smaller than the
total number of broad line clouds.  First, individual clouds on randomly
oriented Keplarian orbits filling a spherical volume might have a
distribution preferentially in the plane of the disk (Osterbrock~1993 cited
in Dietrich et al.~1999). Second, if the cloud velocity distribution is
dominated by an inflow or an outflow bulk velocity, then the number of
transiting clouds would also be much smaller than the total (because the
radial component of the velocity is much larger than the tangential
component). The velocity dispersion of the clouds can be inferred 
from the line widths,
assuming a particular geometry.  In this paper we adopt the simplest model,
assuming that the velocity distribution of the occulting clouds is randomly
oriented in a spherical shell and is induced by the gravitational force of
the central black hole.  The characteristic distance of the shell from the
central source ($\sim 10^4$ Schwarzschild radii) is then larger by three
orders of magnitude than the scale of the continuum emission region (see,
e.g. Peterson~1997).

We model the lensed optical source as an accretion disk surrounded by a
large number of optically thick, identical spherical clouds. We adopt the
radial surface brightness profile for a thermal accretion disk as computed
by Agol \& Krolik~(1999).  The broad line clouds are assumed to be located
in a spherical shell of radius $R$ and have orbits with random inclinations
around a black hole of mass $M_{\rm bh}$.  Based on the virial theorem for
the cloud distribution, we get
\begin{equation}
R = \frac{r_{\rm sch}}{2f_{\rm g}}\left(\frac{c}{\sigma_{\rm cl}}\right)^2,
\end{equation}
where $\sigma_{\rm cl}$ is the 1-D velocity dispersion of the broad lines
in ${\rm km\,sec^{-1}}$, $r_{\rm sch}=2GM_{\rm bh}/c^2$ is the
Schwarzschild radius of the black hole, $G$ is the gravitational constant,
and $c$ the speed of light. The constant $f_{\rm g}$ depends the geometry of the
cloud velocity distribution; in our spherical shell (2-D) geometry $f_{\rm g}=2$
(an isothermal sphere would be described by $f_{\rm g}=3$). Given a total number 
of clouds $N_{\rm cl}$, the covering fraction is
$F_{\rm c}=N_{\rm cl}\pi r_{\rm cl}^2/4\pi R^2$, where $r_{\rm cl}$
is the physical radius of each cloud. Therefore
\begin{equation}
r_{\rm cl} = R\sqrt{\frac{4F_{\rm c}}{N_{\rm cl}}}
\label{eq:second}
\end{equation}
Throughout this paper we show numerical examples with typical parameter
values of $M_{\rm bh}=5\times10^8M_{\odot}$, $\sigma_{\rm cl}=5\times
10^3\,{\rm km\,sec^{-1}}$ and $F_{\rm c}=0.1$, and assume a face-on disk
profile observed in the $R$-band. We show results as a function of the
number of clouds, $N_{\rm cl}$, which is the free parameter that is least
constrained by existing observations of quasars.
 
\section{Microlensing Model}
\label{mlens}

Throughout most of this paper we consider a fiducial lensing scenario,
comprising a lens galaxy at a redshift $z_{\rm d}=0.5$ and a source at
$z_{\rm s}=1.5$. We assume typical microlensing parameters encountered just
inside the Einstein radius of the lens galaxy, which is modeled as a
spherical singular isothermal sphere. In particular, we adopt the likely
values of $\kappa_\star=0.08$ and $\kappa_{\rm c}=0.46$ for the convergence
in stars and smoothly distributed mass, respectively, and a value of
$\gamma=0.54$ for the shear (see Wyithe \& Turner~2002). 
The average magnification of the corresponding macro-image is
12.5. The magnification maps\footnote{The average number of rays per pixel
in our magnification maps is greater than $N_{\rm rays}=100$. Thus, the maps
are computed to an accuracy better than $N_{\rm rays}^{0.25}/N_{\rm
rays}\sim3\%$ per pixel, and a significantly higher accuracy for high
magnification pixels. Note that the scaling is with $N_{\rm rays}^{0.25}$
rather than the Poisson scaling of $N_{\rm rays}^{0.5}$ because rays are
placed on a regular grid in the image plane rather than being randomly
distributed.  Careful examination of ``light-curves'' for individual
clouds clearly demonstrates that the variability seen in Figure~\ref{fig2} 
is not caused by simulation noise.} for the simulations in this paper
were computed with the \textit{microlens} program kindly provided to us by
Joachim Wambsganss.

\section{Variability due to Obscuration of the Disk by Broad Line Clouds}
\label{BLR}

Monitoring data of the the systems RX~J0911+05 and SBS~1520+530 by Burud~(2002) 
shows evidence for short-term variability which is unique to individual images. 
The variability is observed on time-scales of tens to hundreds of days, with an 
amplitude of up to a few percent. Furthermore, the microlensing features appear
 to be associated with small intrinsic fluctuations. In this section we present
a microlensing model that accounts for all three of these features.

\subsection{Intrinsic and Microlensing Lightcurves} 

\begin{figure*}[hptb]
\epsscale{.7}
\plotone{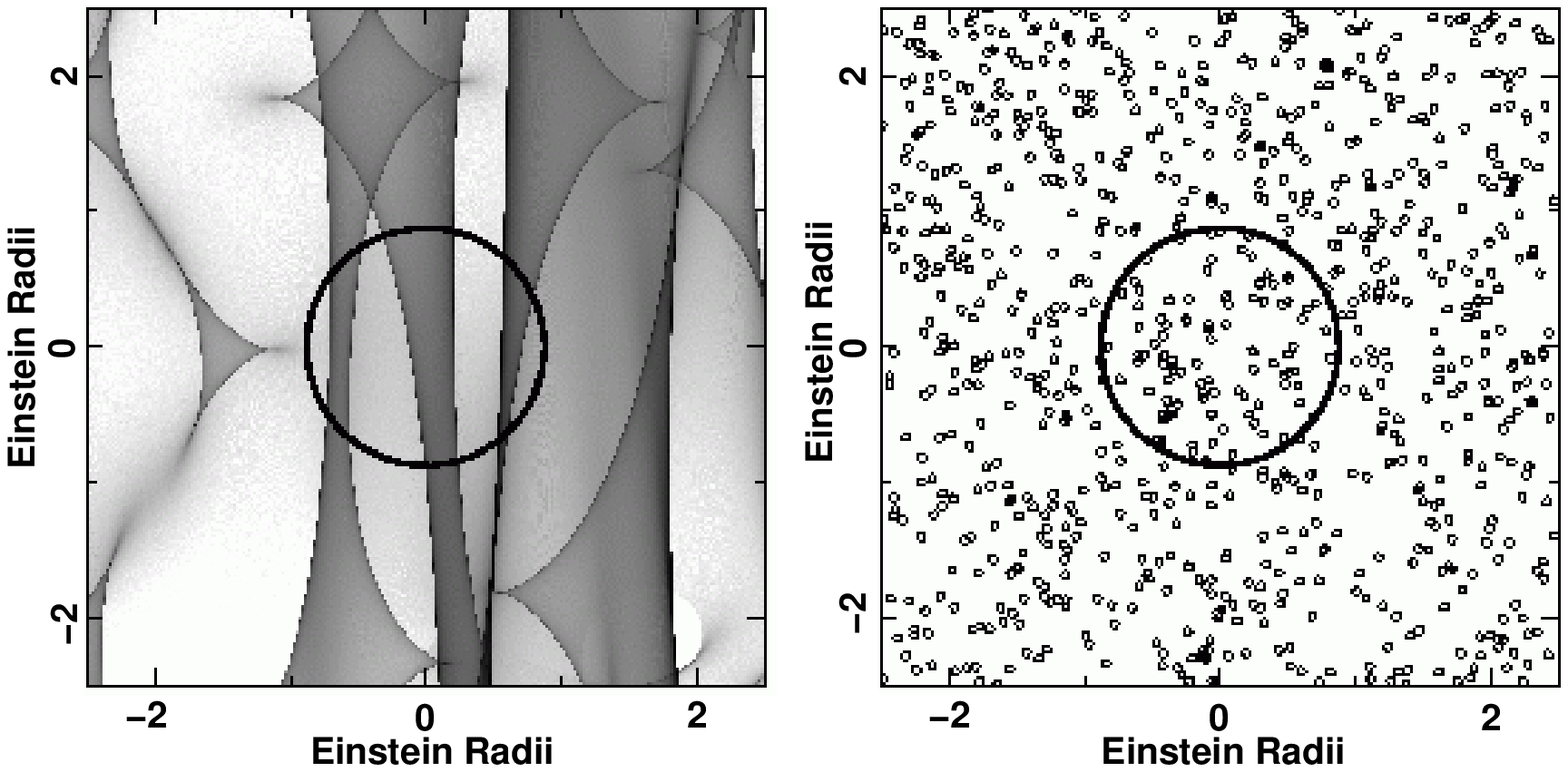}
\caption{\label{fig1} The microlensing geometry. {\it Left}: A microlensing
magnification map with the contour enclosing 95\% of the flux from a
face-on thermal accretion disk (Agol \& Krolik~1999) around a
$5\times10^8M_{\odot}$ black hole. {\it Right}: The projection of a shell of
$10^5$ randomly distributed clouds assuming a covering factor of 10\% and a
velocity dispersion of 5000 ${\rm km\,s\,^{-1}}$. The lens and source
redshifts are $z_{\rm d}=0.5$ and $z_{\rm s}=1.5$. The microlens masses are
$m=0.1M_{\odot}$.}
\end{figure*}

Figure~\ref{fig1} shows the geometry of the system. The left panel shows a
portion of a magnification map computed for the aforementioned parameter
values. Superimposed on this map is a contour enclosing 95\% of the flux
from the accretion disk source. This contour is displayed again on the
right hand panel of Figure~\ref{fig1}, together with a random distribution
of broad line clouds, whose sizes were computed for $N_{\rm cl}=10^5$ [see
Eq.~(\ref{eq:second})].  Since the velocities of the broad line clouds are
an order of magnitude larger than the projected transverse velocity expected
for the
lens galaxy, we assume the accretion disk to be stationary relative to the
caustic network. The clouds obscure a fraction $F_{\rm c}=0.1$ of the disk
surface. Since the disk surface brightness is a function of radius, the
motion of the clouds across regions of varying surface brightness causes low 
level variability of the total flux. Variability also results
from the Poisson noise associated with the variance in the number of
obscuring clouds.  In the absence of microlensing, and assuming a stationary 
disk the intrinsic observed flux as a function of time, $t$, is
\begin{equation}
f_{\rm int}(t) = \int_0^\infty r\,dr\,s_\nu (r) -
\sum_{i=1}^{N_{\rm cl}}\int_0^{r_{\rm cl}} r\,dr \int_0^{2\pi} d\theta ~ s_\nu
\left(\left|\vec{r_i}(\frac{t}{1+z_{\rm s}}) + \vec{r}\right|\right),
\end{equation} 
where $s_\nu(r)$ is the surface brightness of the disk at frequency $\nu$
and radius $r=|\vec{r}|$, and $\vec{r_i}=(r_i,\theta_i)$ are the time
dependent coordinates of the $N_{\rm cl}$ broad line clouds. 
If the accretion disk is also subject to
microlensing, then the microlensed surface brightness profile is the
product of the position-dependent magnification $\mu(\vec{r})$ with the
intrinsic accretion disk brightness $s_\nu(|\vec{r}|)$.  Thus, in the
presence of microlensing the light-curve is
\begin{eqnarray}
\nonumber &&f_{\rm ml}(t) = \int_0^\infty r\,dr \int_0^{2\pi}
d\theta\mu(\vec{r}) s_\nu (|\vec{r}|) \\ &&- \sum_{i=1}^{N_{\rm
cl}}\int_0^{r_{\rm cl}} r\,dr \int_0^{2\pi} d\theta
\mu\left(\vec{r_i}(\frac{t}{1+z_{\rm s}}) +
\vec{r}\right)s_\nu\left(\left|\vec{r_i}(\frac{t}{1+z_{\rm s}}) +
\vec{r}\right|\right).
\end{eqnarray} 
The first term on the right-hand-side of this equation (the magnified flux
from the disk) is not a function of time since we have assumed a stationary
disk.  The variability in the continuum emission due to obscuration by 
broad line clouds is very different from the
variability in the line profiles due to microlensing of the broad line
emission itself (Schneider \& Wambsganss~1990).

\begin{figure*}[hptb]
\epsscale{.7}
\plotone{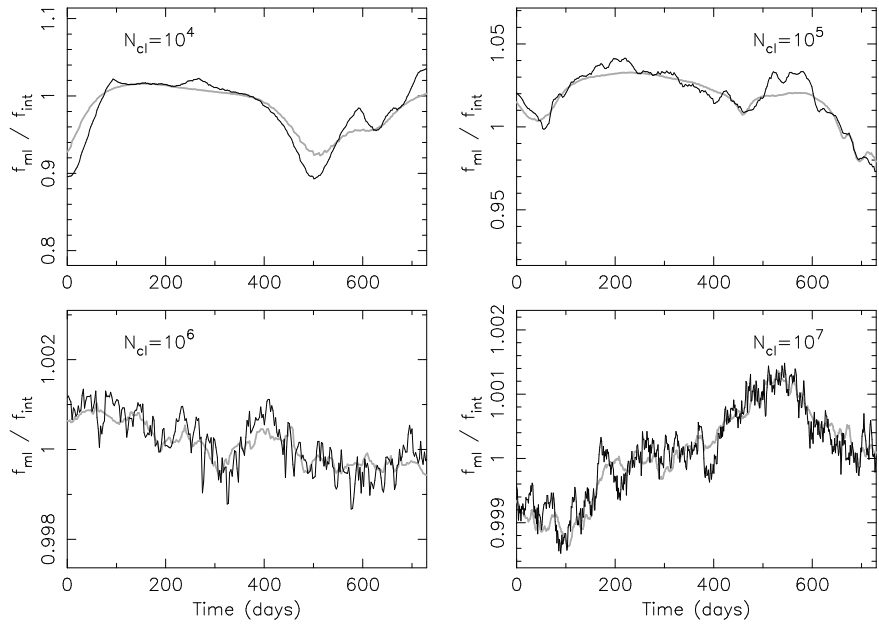}
\caption{\label{fig2} Sample light curves for a disk centered on the
position shown in Figure~\ref{fig1}. Curves are shown assuming four
different cloud sizes corresponding to a total number of clouds per $4\pi$
steradian of $N_{\rm cl}= 10^4$, $10^5$, $10^6$ and $10^7$. The light line
shows the variability in the absence of microlensing, and the dark line
shows the variability with microlensing. Note that the vertical axis has a
different scale in each case. The lens and source redshifts are $z_{\rm
d}=0.5$ and $z_{\rm s}=1.5$.}
\end{figure*}

In Figure~\ref{fig2} we show four sample light-curves corresponding to
$N_{\rm cl}=10^4$, $10^5$, $10^6$ and $10^7$, all with a duration of two
years. The light line shows the level of intrinsic variability $f_{\rm
int}(t)$, while the dark line shows variability including microlensing
$f_{\rm ml}(t)$. It is apparent that larger clouds produce variability of
longer duration and larger amplitude.  The amplitudes range from a few
percent for the largest clouds under consideration down to hundredths of a
percent for the smallest clouds (note that the scaling of the $y$-axis is
different in each panel of Figure~\ref{fig2}). Similarly, the time-scales
for the variability range between tens and hundreds of days. The two 
curves show similar overall trends, however the peaks and troughs in the
microlensed light-curve are more pronounced.  The difference between
the microlensed and intrinsic variability arise because the variation in
effective surface brightness over the accretion disk surface is more
extreme in the microlensed case. Keeping in mind the simplicity of our
model, the light-curves computed using $N_{\rm cl}\sim10^5$ in 
Figure~\ref{fig2} show a striking qualitative resemblance to the light-curves
presented by Burud~(2002), with anomalous microlensing variability superimposed
on, and concurrent with, intrinsic light-curve features.

The variability shown in Figure~\ref{fig2} (both $f_{\rm int}$ and $f_{\rm ml}$) 
is superimposed upon additional
 longer time-scale intrinsic source variability (common to all of its images),
which is observed in unlensed quasars (e.g. Webb \& Malkan~2000).  
We have assumed a population of identical clouds. However for a
real quasar we expect a spectrum of cloud sizes with a distribution
$\phi(r_{\rm cl})$, representing the number of clouds with radii between
$r_{\rm cl}$ and $r_{\rm cl}+\Delta r_{\rm cl}$.  Hence, the light-curves
reflect a superposition of the corresponding amplitudes and time-scales due
to this distribution. The total number of clouds is $N_{\rm cl}^{\rm
tot}=\int_0^\infty dr_{\rm cl}\,\phi(r_{\rm cl})$. However, the effective
number of clouds that contribute to the microlensing signal ($N_{\rm eff}$)
corresponds to an effective cloud radius $r_{\rm cl}^{\rm eff}$ calculated
from the mean of the obscuration area weighted cloud-size distribution. For
spherical clouds
\begin{equation}
N_{\rm eff} \sim 4F_{\rm c}\left(\frac{R}{r^{\rm eff}_{\rm cl}}\right)^2
\hspace{5mm}{\rm where}\hspace{5mm}(r^{\rm eff}_{\rm cl})^2 =
\frac{\int_0^\infty dr_{\rm cl}\,r_{\rm cl}^4\phi(r_{\rm
cl})}{\int_0^\infty dr_{\rm cl}\,r_{\rm cl}^2\phi(r_{\rm cl})}.
\end{equation}
The values of $N_{\rm cl}$ quoted in this paper should be identified with
$N_{\rm eff}$.

\subsection{Variability Statistics}

\begin{figure*}[hptb]
\epsscale{.5} \plotone{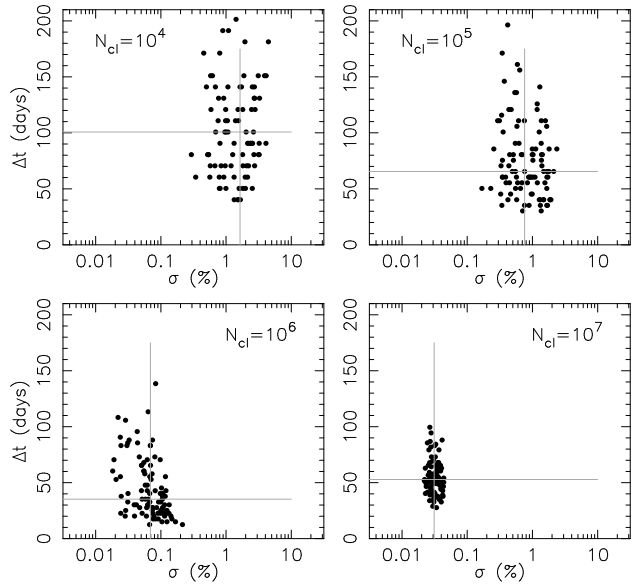}
\caption{\label{fig3} Scatter plots of the correlation time-scale $\Delta
t_{\rm corr}$ versus the variance of the variability amplitude $\sigma$ for
different source positions on the magnification map. The light lines show
the medians of both variables.  The four panels correspond to $N_{\rm cl}=
10^4$, $10^5$, $10^6$ and $10^7$. The lens and source redshifts are fixed
at $z_{\rm d}=0.5$ and $z_{\rm s}=1.5$.  }

\end{figure*}

\begin{figure*}[hptb]
\epsscale{.5}
\plotone{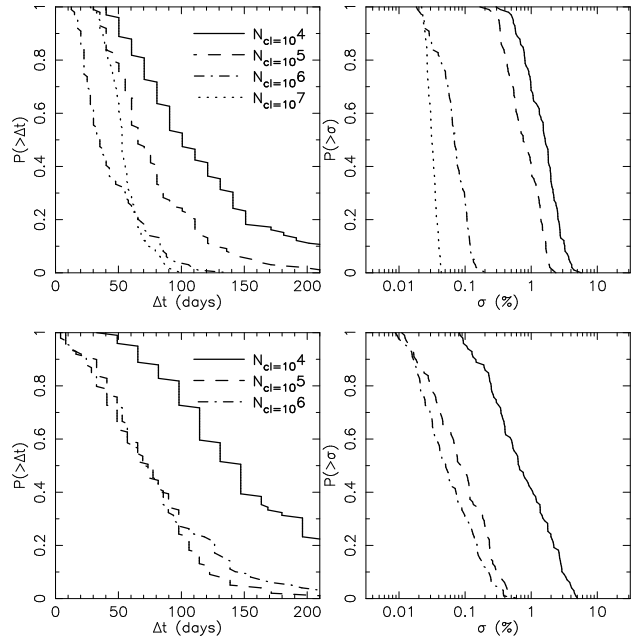}
\caption{\label{fig4} One minus the cumulative probabilities for the correlation
time-scale $\Delta t_{\rm corr}$ (left) and the variability variance
$\sigma$ (right) for the fiducial time-delay lens case (top, $z_{\rm
d}=0.5$ and $z_{\rm s}=1.5$) and for a lensing configuration similar to
Q2237+0305 (bottom, $z_{\rm d}=0.05$ and $z_{\rm s}=1.5$). Curves are shown
for $N_{\rm cl}= 10^4$, $10^5$, $10^6$ and $10^7$ (top only).}
\end{figure*}

The light-curves of Figure~\ref{fig2} demonstrate that bigger clouds
produce variability of a longer duration and a larger amplitude. As can be
seen from Figure~\ref{fig1}, these light-curves were computed (for the
purpose of demonstration) using a favorable source location on the
magnification map close to several caustics. In this 
section we present the characteristic
time-scales and variability amplitudes for 100 random accretion disk
positions across the network of microlensing caustics. Since intrinsic
variability will be present in all images of a multiply imaged quasar and
only microlensing can produce the anomalous light-curve variability, we
calculate variability statistics for the ratio, $f(t)=f_{\rm int}(t)/
f_{\rm ml}(t)$ between the intrinsic and microlensed light-curves at each
source position. We calculate the autocorrelation function, $f_{\rm
AC}(\Delta t) = {\rm sign}(f_{\rm AC}')\sqrt{|f_{\rm AC}'|}$, where $f_{\rm
AC}' \equiv \langle (f(t)-\langle f\rangle)\times(f(t+\Delta t)-\langle
f\rangle)\rangle$. Here, angular brackets denote averaging over long
times. The characteristic variability amplitude is taken to be
$\sigma=f_{\rm AC}(0)$ (the light-curve variance), and the correlation
time-scale $\Delta t_{\rm corr}$ is defined by the condition $f_{\rm
AC}(\Delta t_{\rm corr})/f_{\rm AC}(0)=0.5$ for each source
position. Figure~\ref{fig3} shows scatter plots of $\Delta t_{\rm corr}$
versus $\sigma$ for $N_{\rm cl}=10^4$, $10^5$, $10^6$ and $10^7$. The
aforementioned dependence of the time-scale and variability amplitudes on
$N_{\rm cl}$ are
readily apparent in this plot. The upper panels of Figure~\ref{fig4} show
one minus the single variable cumulative probabilities for $\sigma$ and
$\Delta t_{\rm corr}$. Characteristic time-scales of 50--100 days are
consistent with all cloud sizes, while time-scales below $\sim50$ days are
not consistent with $N_{\rm cl}\la10^4$ and time-scales above $\sim100$
days are only consistent with $N_{\rm cl}\la10^6$. Inspection of the
top-right-hand panel of Figure~\ref{fig4} shows an even stronger dependence
on $N_{\rm cl}$. In particular if $N_{\rm cl}\sim10^4-10^5$, the
variability amplitude is a few tenths of a percent to a few percent, while
the variability amplitude is always below $\sigma\sim0.2\%$ if $N_{\rm
cl}\ga10^6$.

Another heavily monitored gravitational lens is Q2237+0305 at an unusually
low lens redshift of $z_{\rm d}= 0.0394$ (Irwin et al.~1989; Corrigan et
al~1991; $\O$stensen et al.~1995; Wozniak et al.~2000a; Wozniak et
al.~2000b). The lens was monitored not for the purpose of measuring time
delays (for which its geometry is unsuited), but of observing quasar
microlensing (for which it is the most favorable lens known).  The
variability record shows long-term, large amplitude variation, but {\it
no} variation at the percent level over time-scales of tens of days. Thus
if the model proposed in this paper is correct, it must not predict rapid, 
low amplitude
variability in Q2237+0305. To demonstrate the required consistency, we have
repeated the calculation described above for a lens galaxy at $z_{\rm
d}=0.05$ rather than $z_{\rm d}=0.5$ which closely mimics the Q2237+0305
geometry. The lower panel of Figure~\ref{fig4} shows one minus the single
variable cumulative probabilities for $\sigma$ and $\Delta t_{\rm corr}$ as
before for $N_{\rm cl}=10^4$, $10^5$, and $10^6$. The median variability
for $N_{\rm cl}=10^4$ is around $\sigma\sim1\%$ with a range of
$0.1\%\la\sigma\la5\%$. On the other hand, if $N\ga10^5$ the range is
$0.01\%\la\sigma\la0.5\%$, with a median of $\sigma\sim0.1\%$. Thus we
would not expect a previous detection of short-term anomalous variability
in Q2237+0305 if $N_{\rm cl}\ga10^5$. The time-scales are comparable to
those for the generic lens case, which is expected since the timescale 
depends on velocities generated in the source plane rather than the lens plane.

The difference in the statistics of rapid, low amplitude microlensing
variability (due to obscuration of the continuum by broad line clouds)
between the typical lens geometry ($z_{\rm d}\sim0.5$) and the special case
of a lens at very low redshift ($z_{\rm d}\sim0.05$) follows from the
different size of the projected microlens Einstein radius. The
broad-line-cloud induced variability requires that a caustic lie across the
accretion disk. In the typical lensing case ($z_d=0.5$), the source is 
relatively large with respect to the caustic network, so that one or more 
caustics generally cross the source (see Fig.~\ref{fig1}). In the low redshift 
lens case ($z_d=0.05$),
caustic crossings are rare, and so are the broad line microlensing
events. Note that the value of $N_{cl}\sim10^5$ bracketed by the
non-detection at $z_d=0.05$ and the detection at $z_d=0.5$ corresponds to the 
expected number of broad line clouds in the bloated star model (Alexander
\& Netzer~1997).

\section{Viability of Alternative Explanations} 

Two alternative explanations for rapid, low amplitude variability appear in the
literature. We discuss these in turn, and demonstrate their shortcomings in
 explaining the observations of RX~J0911+05 and SBS~1520+530 by Burud~(2002).

\subsection{Variability Due to Disk Hot Spots} 
\label{spots}

Gould \& Miralda-Escud\'{e}~(1997) suggested that the observed rapid,
low-amplitude variability might result from hot spots on the surface of the
disk. This scenario is similar to ours in so far as the
short time-scale results from the large orbital velocities in the source
plane, while the low amplitude results from the fact that only a small
fraction of the total flux is subjected to large amplitude magnification by
microlensing. The main differences from our model are that the hot spot
velocities ($\sim0.2c$) are larger than the cloud velocities due to their 
closer proximity to the
black-hole; and the amplitude of fluctuations is governed by the contrast
between the surface brightness of the hot spots and the disk, rather than between
the opaque clouds and the disk.

We make the simple assumption of circular Keplarian rotation for the hot
spots, and distribute a number $N_{\rm sp}$ of them at random within the
contour containing 99\% of the flux from the disk surface. We assume that 
the hot spots are
long lived for the purpose of the light-curve computation. This is
equivalent to the computation of the statistics for the time averaged
number of spots. We denote the radius of each spot by $r_{\rm sp}$, and the
ratio between the surface brightness of the spot and the disk (locally) by
the contrast $C$. Hence, $C=1$ if the spots have the same temperature as
the disk, $C=0$ if the spots have zero temperature and contribute no flux
(equivalent to the obscuration case presented in the previous section), and
$C>1$ for spots that are hotter than the disk.  The microlensing
light-curve for a stationary disk with hot spots is
\begin{eqnarray}
\nonumber &&f_{\rm ml}(t)= \int_0^\infty r\,dr \int_0^{2\pi}
d\theta\mu(\vec{r}) s_\nu (|\vec{r}|)\\ && + (C-1)\sum_{i=1}^{N_{\rm
sp}}\int_0^{r_{\rm sp}} r\,dr \int_0^{2\pi} d\theta
\mu\left(\vec{r_i}(\frac{t}{1+z_{\rm s}}) +
\vec{r}\right)s_\nu\left(\left|\vec{r_i}(\frac{t}{1+z_{\rm s}}) +
\vec{r}\right|\right),
\end{eqnarray} 
where $\vec{r_i}=(r_i,\theta_i)$ are the time
dependent coordinates of the $N_{\rm sp}$ hot spots. This formulation 
does not result in intrinsic variability.

The parameters for this model are not constrained by observations and we 
have computed
statistics for six representative cases. We assume values for $r_{\rm sp}$
of $10^{14}$cm and $5\times10^{14}~{\rm cm}$, with contrasts of $C=50$ and
$C=2$ respectively, resulting in fluctuation amplitudes of
$\sim0.1-1\%$. For each case we compute light-curves for $N_{\rm sp}= 1$,
10 and 100. The corresponding fractions of disk covered by the spots are
$F_{\rm sp}=0.0001$, 0.001 and 0.01 if $r_{\rm sp}=10^{14}$cm and $F_{\rm
sp}=0.0025$, 0.025 and 0.25 if $r_{\rm sp}=5\times10^{14}$cm. The amplitude
of the light-curve scales with $|1+(C-1)F_{\rm sp}|$ and so the results in
this section for the amplitude of the fluctuations can be easily
generalized to other values of $C$.

\begin{figure*}[hptb]
\epsscale{.8}
\plotone{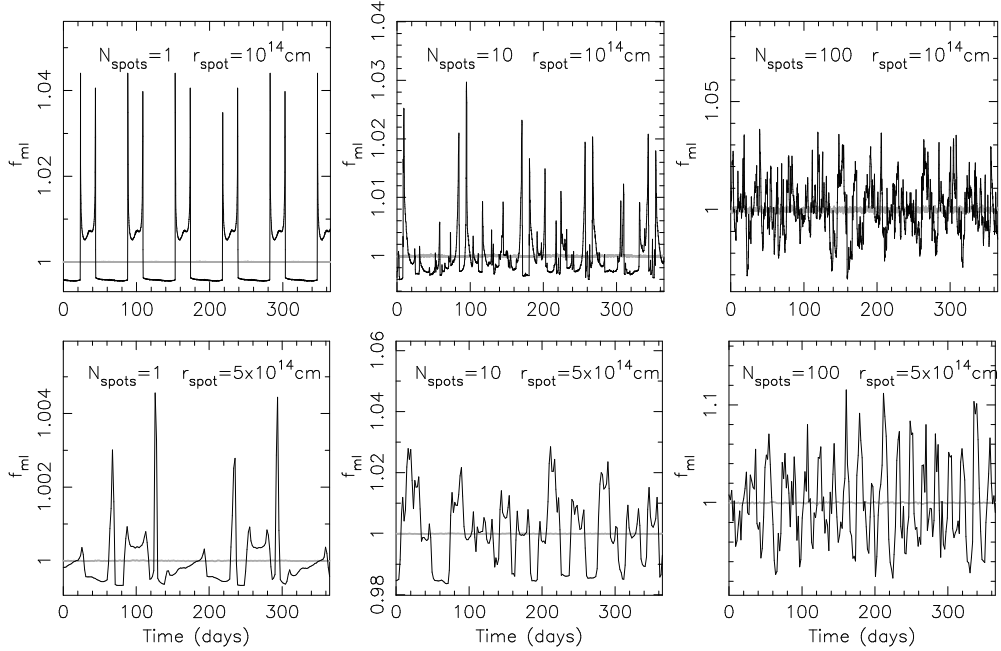}
\caption{\label{fig5} Sample light curves assuming spot radii of $r_{\rm
sp}=10^{14}$cm (upper row) and $r_{\rm sp}=5\times10^{14}$cm (lower
row). In each case, light curves resulting from $N_{\rm sp}=1$, 10 and 100
are shown. The light line shows the lack of variability in the absence of
microlensing, and the dark line shows the variability due to
microlensing. Note that the vertical axis has a different scale in each
case. The lens and source redshifts are $z_{\rm d}=0.5$ and $z_{\rm
s}=1.5$. The contrast of individual spots in the upper and lower rows is
$C=50$ and $C=2$, respectively.}
\end{figure*}

Figure~\ref{fig5} shows sample light-curves for these six cases. The
signature of periodicity noted by Gould \& Miralda-Escud\'{e}~(1997) is
apparent in the light-curves with a single spot, although for $N_{\rm sp}>1$
the variability is due to the sum of many periodic light-curves with a 
random phase and the periodicity is diluted.
Figure~\ref{fig5} implies that the variability time-scale is shorter than
that generated by broad line clouds. Furthermore, if only a few spots are
present then the light-curve resembles a classical microlensing light-curve
with M-shaped events, but with a reduced amplitude and time-scale. Thus,
small numbers of hot spots produce microlensing peaks which look
qualitatively different from the troughs produced in the light-curves by
the broad line clouds. Moreover, in our formulation, hot spots do not 
produce intrinsic variability, so that we do not expect microlensing peaks
to be concurrent with intrinsic peaks as predicted by the broad line
cloud model and observed by Burud~(2002).

\begin{figure*}[hptb]
\epsscale{.8} \plotone{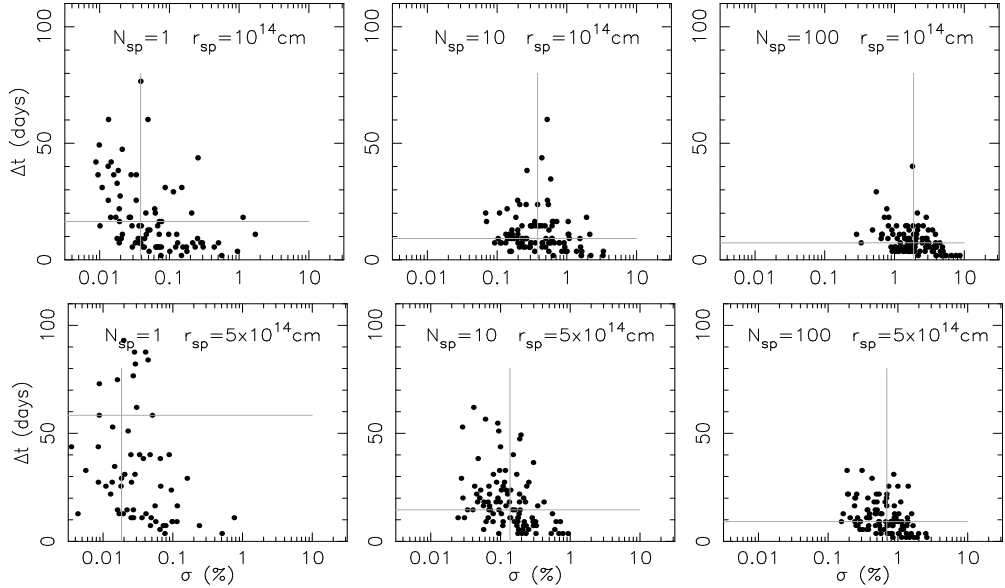}
\caption{\label{fig6} Scatter plots of the correlation time-scale $\Delta
t_{\rm corr}$ versus the variance of the variability amplitude $\sigma$ for
different source positions. The light lines show the medians of both
variables.  The upper and lower rows assume spot radii of $r_{\rm
sp}=10^{14}$cm and $r_{\rm sp}=5\times10^{14}$cm. Plots for $N_{\rm sp}=1$,
10 and 100 are shown in each case. The contrast of individual spots in the
upper and lower rows is $C=50$ and $C=2$, respectively.}
\end{figure*}

Figure~\ref{fig6} quantifies the variability statistics by showing scatter
plots of $\Delta t_{\rm corr}$ versus $\sigma$ for the six cases mentioned
above. The variability amplitudes cannot be directly used to constrain
$N_{\rm sp}$ or $r_{\rm sp}$ since it is governed by the contrast $C$.
However, the time-scales are much shorter than those due to the broad line
cloud absorption, with medians below 10--20 days in all cases except for a
single large spot ($r_{\rm sp}\sim15r_{\rm sch}$). Note that this
time-scale is nearly independent of the black-hole mass; although a smaller
$M_{\rm bh}$ lowers the orbital velocity at a fixed radius, the radius
corresponding to a fixed level of disk emission is roughly proportional to
$M_{\rm bh}$.  Overall, the predicted variability time-scales are shorter than 
those observed in RX J0911+05 and SBS 1520+530.

\subsection{Microlensing by Very Small Masses} 
\label{lcurve}

\begin{figure*}[hptb]
\epsscale{.4}
\plotone{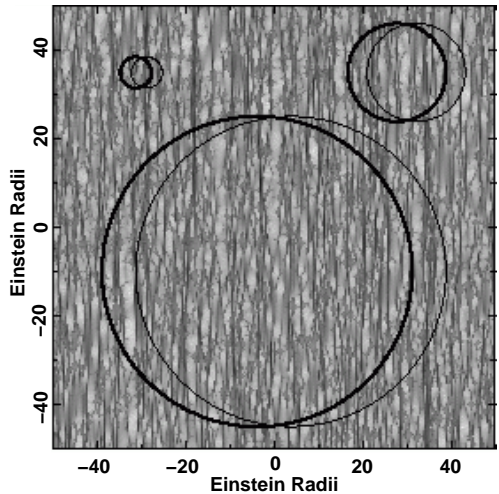}
\caption{\label{fig7} Sample magnification map for very low mass
microlenses.  The contours enclosing 95\% of the flux from a face-on
thermal accretion disk (Agol \& Krolik~1999) around a
$5\times10^8M_{\odot}$ black hole are shown for three microlens masses,
namely $m=10^{-2}M_{\odot}, 10^{-3}M_{\odot}$, and $10^{-4}M_{\odot}$.  The
pair of circles in each case are shown to demonstrate how far the disk
moves during 10 years, assuming a transverse velocity for the lens galaxy
of $400\,{\rm km\,sec^{-1}}$. The lens and source redshifts are $z_{\rm
d}=0.5$ and $z_{\rm s}=1.5$.}
\end{figure*}

\begin{figure*}[hptb]
\epsscale{.7}
\plotone{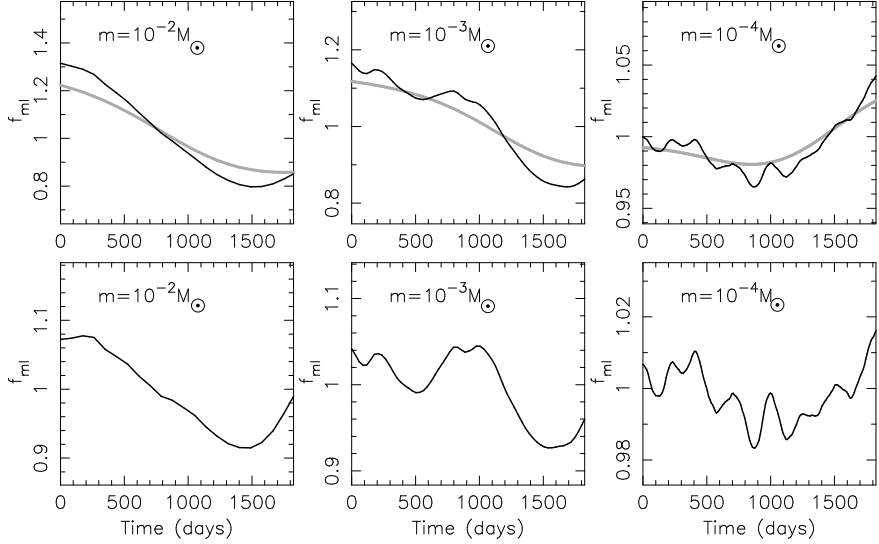}
\caption{\label{fig8} Sample light curves for a smooth accretion disk and
very low mass microlenses.  Curves are shown for the three microlens
masses, $m=10^{-2}M_{\odot}, 10^{-3}M_{\odot}$ and $10^{-4}M_{\odot}$. In
the upper panels the dark lines show raw microlensing light-curves, while
the light lines show the light-curves after being Gaussian-smoothed on the
correlation time-scale. In the lower panels, the curve shows the ratio
between the raw and smoothed curves. The lens and source redshifts are
$z_{\rm d}=0.5$ and $z_{\rm s}=1.5$, and the transverse velocity of the lens
galaxy is $400\,{\rm km\,sec^{-1}}$.}
\end{figure*}

We also consider rapid low amplitude microlensing
variability due to a population of very low mass compact objects. In this
scenario, small amplitude variability results from the source being large
compared to the characteristic scale of the caustic network (or
equivalently the microlens Einstein radius), while the rapid time-scale
results from the short crossing time of this network. We assume a
featureless accretion disk without hot spots around a
$5\times10^8M_{\odot}$ black hole that moves relative to the caustic
network. The transverse velocity $\vec{v}_{\rm gal}$ of the lens galaxy with 
respect to the observer--source line-of sight is assumed to have a magnitude 
of $400\,{\rm km\,sec^{-1}}$. The light-curve $f_{\rm ml}(t)$ is
\begin{equation}
f_{\rm ml}(t)=\int_0^\infty r\,dr \int_0^{2\pi} d\theta\mu\left(\vec{r}_0+ \vec{r}+
(t-t_0)\frac{\vec{v}_{\rm gal}}{1+z_{\rm d}}\frac{D_{\rm s}}{D_{\rm d}}\right)
s_\nu\left(\left|\vec{r}\right|\right),
\end{equation} 
where $\vec{r}_0$ is the quasar position at time $t_0$. $D_d$ and $D_s$ are
the angular diameter distances to the lens and source, which we again take
to have redshifts $z_{\rm d}=0.5$ and $z_{\rm s}=1.5$. A sample
magnification map for very low mass microlenses is shown in
Figure~\ref{fig7}. Superimposed on this map are the contours enclosing 95\%
of the flux from a face-on thermal accretion disk assuming three different
cases of microlens masses, namely $m=10^{-2}M_{\odot}, 10^{-3}M_{\odot}$
and $10^{-4}M_{\odot}$. The pair of circles in each case demonstrates how
far the disk moves during 10 years. Sample light-curves are shown as the
dark lines in the top panel of Figure~\ref{fig8}. The light-curves show two
characteristic time-scales: a long time-scale governed by the caustic
clustering length (see Figure~\ref{fig7}) as well as a shorter
time-scale. 

\begin{figure*}[hptb]
\epsscale{.8} 
\plotone{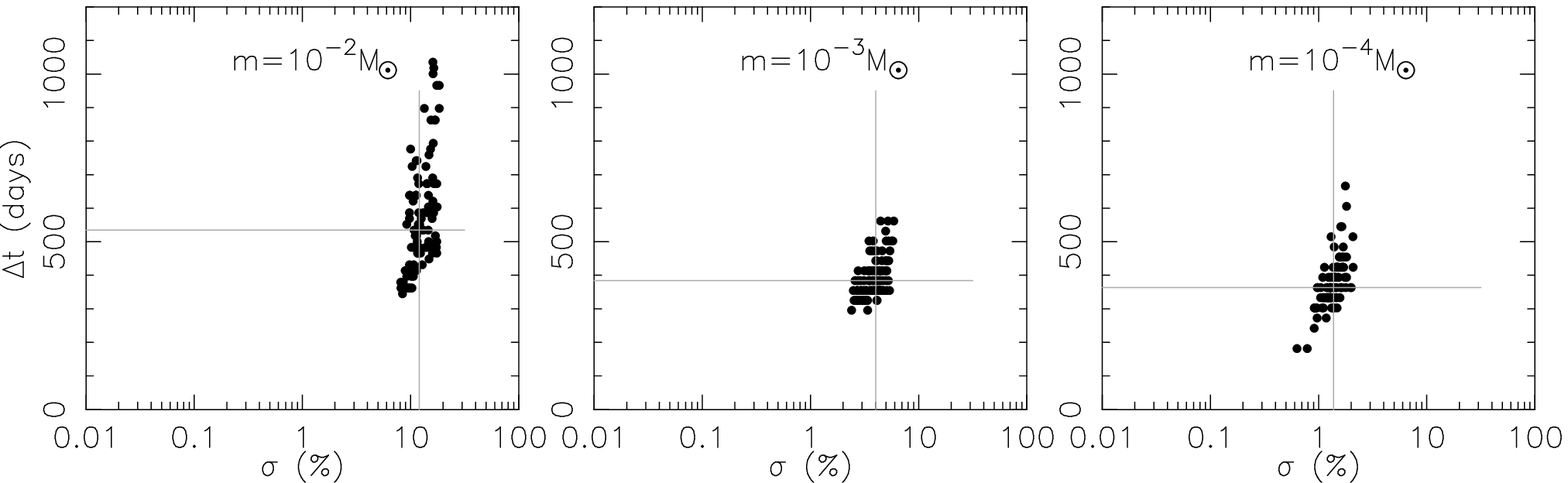}
\caption{\label{fig9} Scatter plots of the correlation time-scale $\Delta
t_{\rm corr}$ versus the variance of the variability amplitude $\sigma$ for
planetary-mass microlenses. The light lines show the medians of both
variables. Plots are shown for microlens masses of $m=10^{-2}M_{\odot},
10^{-3}M_{\odot}$ and $10^{-4}M_{\odot}$.}
\end{figure*}

In order to isolate the characteristic amplitude and time-scales of the 
rapid variability, we performed the following procedure. For each 
light-curve we find the correlation time as before, and then smooth
the light-curve using the correlation time as the standard deviation of a Gaussian
smoothing function.  The resulting curves are shown by the light lines in
the upper panels of Figure~\ref{fig8}. The ratio between the two curves
yields the rapid, low amplitude variability. These are shown in the lower
panels of Figure~\ref{fig8}.  We constructed autocorrelation functions
for a hundred light-curves in each case and extracted the characteristic
variability and amplitude as before. The results are shown in
Figure~\ref{fig9}. We find that the smallest masses under consideration can
produce the required variability amplitudes of $\sim1\%$.  However the
time-scales are around 400 days, longer than those seen in RX J0911+05 and
SBS 1520+530.  Note that $\Delta t_{\rm corr}$ is proportional to the
inverse of the transverse velocity assumed. The time-scale is also
dependent on the direction of the transverse velocity with respect to the
microlensing shear. We have conservatively assumed that the direction to be
parallel to the shear which results in the most rapid time-scales.  However
a transverse velocity parallel to the shear can significantly increase
$\Delta t_{\rm corr}$ (Wambsganss, Paczynski \& Katz~1990; Lewis \&
Irwin~1996) and make it even less compatible with the data. In addition to
the relatively large time-scale for the rapid component of variability, 
we note that the simulated light-curves differ qualitatively from those
observed. The observed light-curves do not show long term microlensing 
variability over the duration of the monitoring period, as is 
predicted by this model. Furthermore, since the microlensing variability 
must be uncorrelated between macro-images, 
there is no mechanism to explain the correlation between the intrinsic variability
and the microlensing variability observed in the light-curves of 
RX J0911+05 and SBS 1520+530. 

To save this model, one might suppose that hot spots are
present in addition to the very low mass microlenses. In this way, one
might shorten the time-scale predicted by the low mass-microlensing
light-curves using the large source plane velocities of the hot-spots, 
while retaining the low amplitudes for the
reasons discussed in \S~\ref{spots}. However we argue that this scenario
will also be inconsistent with the observations. As can be seen from
Figure~\ref{fig7}, the short ($\sim400$ day) variability is superimposed on
longer term variability that was removed in the above calculations of
variability statistics. This long term variability, which is due to the
clustering of caustics on scales many Einstein Radii in extent will still
be present following the addition of hot spots, but is not seen in the 
light-curves of RX J0911+05 and SBS 1520+530, which have
durations of $\sim1000-1500$ days. Thus
predictions of microlensing variability due to very low mass microlenses
differ qualitatively from observations, whether or not the disk has spots.

\section{Discussion}
\label{discussion}

We have identified a simple explanation for the anomalous microlensing
variability reported by Burud~(2002) in the systems RX J0911+05 and SBS
1520+530. We find that the obscuration of a differentially magnified
(microlensed) accretion disk by optically-thick broad-line clouds results
in rapid variability due to the high cloud velocities and in fluctuations 
with a low amplitude due to the large number of clouds (and hence small level of
Poisson fluctuations).  The model predicts fluctuations with an amplitudes
(a few percent) and time-scales (50-100 days), comparable to those observed. 
In addition, the model naturally explains the observed correlation between 
the intrinsic and microlensing features. The model does not require the 
inclusion of any hypothetical components.

It is in principle also possible to generate rapid, low amplitude variability through
microlensing of hot spots on the surface of an accretion disk (Gould \& 
Miralda-Escud\'{e}~1997) or through very low mass microlenses. In the first 
case, our simulations show that the
variability time-scales are substantially shorter than those observed in RX
J0911+05 and SBS 1520+530.  Of the cases considered, the longer time-scales
are produced by a small number of relatively large spots ($r_{\rm sp}\sim10
r_{\rm sch}$). However, a small number of spots produces light-curve shapes
that differ qualitatively from those identified in the observations. In the
second alternative model of planetary-mass micolenses, we find that the
time-scales are governed by the caustic clustering length and source crossing
time, rather than the crossing time of the microlens Einstein radius. As
a result, the predicted time-scales are longer than those observed.

If our explanation for the nature of this unexpected microlensing
signal is correct, it will be of great significance in constraining 
the properties of the broad line region. Although the dimension 
of the broad line region is measured
by reverberation mapping and the covering factor is known to be
$\sim10\%$, there is currently little information regarding the number and
hence the size, of individual broad line clouds (see discussion in \S
\ref{quasar}). Our simple model suggests that to explain the anomalous
light-curve features observed by Burud~(2002) in RX J0911+05 and SBS
1520+530, the number of broad line clouds that contribute the bulk of the
covering factor should be $N_{\rm cl}<10^6$. In addition, the lack of these
features in Q2237+0305 suggests that $N_{\rm cl}>10^4$. Interestingly,
these constraints bracket the predicted range for the number of clouds in
the bloated star model (Alexander \& Netzer 1997).

Our results provide a qualitative explanation for the rapid, low amplitude
microlensing variability recently observed in several gravitational lens
systems. As more measurements of variability become available in the future
for these and other lens system, it will become possible to refine our
analysis by detailed modeling of individual systems. The sub-microarcsecond
resolution provided by the sensitivity of the rapid microlensing
fluctuations to the number of broad line clouds, has the potential to place
important constraints on models of the broad line region.

\acknowledgements 

We greatfully acknowledge the use of the model accretion disk calculated by
Eric Agol and the \textit{microlens} program written by Joachim
Wambsganss, as well as helpful comments from Josh Winn, Rachel Webster and 
Hagai Netzer. 
This work was supported in part by NASA grants NAG 5-7039,
5-7768, and NSF grants AST-9900877, AST-0071019 for AL. JSBW is supported
by a Hubble Fellowship grant from the Space Telescope Science Institute,
which is operated by the Association of Universities for Research in
Astronomy, Inc., under NASA contract NAS 5-26555.

\vspace{30mm}

\begin{thebibliography}{}

\bibitem[]{}
Agol, E., \& Krolik, J. 1999, ApJ, 524, 49

\bibitem[]{}
Alcock, C., et al., 2000, 542, 281

\bibitem[]{}
Alexander, T., Netzer, H., 1997, MNRAS, 284, 967

\bibitem[Arav et al.(1998)]{1998MNRAS.297..990A} Arav, N., Barlow, T.~A.,
Laor, A., Sargent, W.~L.~W., \& Blandford, R.~D.\ 1998, \mnras, 297, 990.

\bibitem[]{}
Burud, I., et al., 2000, ApJ., 544, 117

\bibitem[]{} Burud, I., 2002, PhD Thesis, Institut d'Astrophysique,
Liege,\newline
http://vela.astro.ulg.ac.be/themes/dataproc/deconv/theses/burud/ibthese\_e.html

\bibitem[]{}
Corrigan et al., 1991, Astron. J., 102, 34

\bibitem[]{}
Dietrich, M., Wagner, S.J., Courvoisier, T.J.-L., Bock, H., North, P., 1999, 
Astron. Astrophys., 351, 31 

\bibitem[]{}
Gould, A., Miralda-Escud\'{e}, J., 1997, ApJ., 483, L13

\bibitem[]{} Irwin, M. J., Webster, R. L., Hewitt, P. C., Corrigan, R. T.,
Jedrzejewski, R. I., 1989, Astron. J., 98, 1989
 
\bibitem[]{}
Kochanek, C.S., 2002, ApJ., submitted, astro-ph/0204043

\bibitem[]{}
Krolik, J.H., McKee, C.F., Tarter, C.B., 1981, ApJ, 249, 422

\bibitem[]{}
Kundic, T., et al., 1997, ApJ, 482, 75

\bibitem[]{}
Lewis, G.F., Irwin, M.J., 1996, MNRAS, 283, 225 

\bibitem[]{}
Netzer, H., 1990, in Blandford, R.D., Netzer, H., Woltjer, L., eds, Saas-Fee Advanced course 20: Active galactic Nuclei. Springer, NewYork, p. 57

\bibitem[]{}
$\O$stensen, R. et al. 1996, Astron. Astrophys., 309, 59

\bibitem[]{}
Osterbrock D.E., 1993, ApJ 404, 551

\bibitem[]{}
Peterson, B.M., 1997, An introduction to Active Galactic Nuclei, Cambridge 
University Press                        

\bibitem[]{}
Refsdal, S., 1964, MNRAS, 128, 307

\bibitem[]{}
Refsdal, S., Stabell, R., Pelt, J., Schild, R., 2000, Astron. Astrophys., 360, 10

\bibitem[]{}
Schechter, P.L., et al. 1997, ApJ, 475, L85

\bibitem[]{}
Schild, R., 1996, ApJ, 464, 125

\bibitem[]{}
Schneider, P., Wambsganss, J., 1990, Astron. Astrophys, 237, 42

\bibitem[]{}
Webb, W., Malkan, M., 2000, ApJ., 540, 652

\bibitem[]{} Wozniak, P. R., Alard, C., Udalski, A., Szmanski, M., Kubiak,
M, Pietrzynski, G., Zebrun, K., 2000a, Ap. J., 529, 88

\bibitem[]{} Wozniak, P. R., Alard, C., Udalski, A., Szmanski, M., Kubiak,
M, Pietrzynski, G., Zebrun, K., 2000b, Ap. J., 540, L65

\bibitem[]{}
Wambsganss, J., Paczynski, B., Katz, N., 1990, ApJ., 352, 407

\bibitem[]{}
Wambsganss, J., Schmidt, R.W., Colley, W.N., Kundic, T., Turner, E.L., 
2000, Astron. Astrophys., 362, L37

\bibitem[]{}
Wyithe, J. S. B, Webster, R. L., Turner, E. L., 2000, MNRAS, 315, 51

\bibitem[]{}
Wyithe, J. S. B, Webster, R. L., Turner, E. L., 2000, MNRAS, 318, 762

\bibitem[]{}
Wyithe, J. S. B, Turner, E. L., 2002, ApJ., accepted, astro-ph/0203214 

\end{thebibliography}
\end{document}